\documentstyle[12pt,epsfig]{article}

\begin{document}
\baselineskip=23pt

\vspace{1.2cm}

\begin{center}
{\Large \bf  Massive Gauge Bosons in Yang-Mills Theory without Higgs
Mechanism}

\bigskip

Xin-Bing Huang\footnote{huangxb@shnu.edu.cn}\\
{\em Shanghai United Center for Astrophysics (SUCA),} \\
{\em   Shanghai Normal University, No.100 Guilin Road, Shanghai
200234, China}
\end{center}

\bigskip
\bigskip
\bigskip

\centerline{\large Abstract}

Two kinds of Yang-Mills fields are found upon the concepts of mass
eigenstate and nonmass eigenstate. The Yang-Mills fields of the
first kind were proposed by Yang and Mills, which couple to the mass
eigenstates with the same rest mass, whose gauge bosons are
massless. I find that there are second kind of Yang-Mills fields,
which are constructed on a five-dimensional manifold. Only the
nonmass eigenstates couple to the Yang-Mills fields of the second
kind, which are the nonmass eigenstates as well and composed of mass
eigenstates of gauge bosons. The mass eigenstates of the Yang-Mills
fields of the second kind live in the four-dimensional spacetime,
the corresponding gauge bosons of which may be massive. The
$SU(2)\times U(1)$ gauge fields of the second kind are studied
carefully, whose gauge bosons, which are the mass eigenstates, are
the $W^{\pm}$, $Z^{0}$ and photon fields. The rest masses of
$W^{\pm}$ and $Z^{0}$ obtained are the same as that given by the
Glashow-Salam-Weinberg model of electroweak interactions. It is
discussed that this model should be renormalizable.

\vspace{1.2cm}


\vspace{1.2cm}

\newpage

55 years ago, Yang and Mills constructed the gauge field theory of
non-Abelian group, which has become the most fundamental content in
quantum field theory. Upon the principle that physical laws should
be covariant under the local isospin rotation they proposed the
$SU(2)$ Yang-Mills theory~\cite{yan54}. But they could not obtain
the massive gauge bosons then. About 10 years later, an ingenious
trick called the Higgs mechanism was independently invented by Higgs
and Englert and Brout~\cite{hig64}, who introduced a scalar field
and the spontaneous symmetry broken mechanism of vacuum by fixing a
vacuum expectation value of the scalar field and make the
intermediate vector bosons obtain masses.

Based on the Yang-Mills fields and the Higgs mechanism, Glashow,
Salam and Weinberg {\em etc.} proposed a renormalizable theory
unifying the weak and electromagnetic interactions, namely
$SU_{L}(2)\times U_{Y}(1)$ gauge theory~\cite{gsw60}. Although this
electroweak theory had predicted the masses of intermediate vector
bosons, which were confirmed by experiments, there are still several
unconfirmed predictions or conflicting phenomena in it. e.g.
Firstly, experimenters have not found any hints of the Higgs boson
till now; Secondly, a lot of recent experiments imply that the
neutrinos should be massive and be mixed~\cite{xin04}. Here I
discuss a model to give the massive gauge bosons in Yang-Mills
theory without Higgs mechanism.

In this letter, the signature of spacetime metric
$\eta_{\mu\nu}(\mu,\nu=0,1,2,3)$ is $(+,-,-,-)$, and the spacetime
coordinates are described by the contravariant four-vector $x^{\mu}$
($\hbar=c=1$ is adopted). In Ref.\cite{hua09a}, the rest mass
operator\footnote{I use $\partial_{z}\equiv\frac{\partial}{\partial
z}~,~\partial_{\mu}\equiv\frac{\partial}{\partial x^{\mu}}$ and
$\partial_{\alpha}\equiv\frac{\partial}{\partial x^{\alpha}}$.}
\begin{equation}
\label{1000} \hat{m}=-i\partial_{z}
\end{equation}
is defined by introducing an extra parameter $z$ besides of the
spacetime coordinates $x^{\mu}$. From the mathematical point of
view, $z$ and $x^{\mu}$ establish a five-dimensional manifold. The
definition of the rest mass operator leads to a theorem that a field
${\cal F}(x,z)$ is massless if and only if ${\cal F}(x,z)$ is
$z$-independent~\cite{hua09a}. Hence the massless gravitational
field, the electromagnetic field and $SU(3)$ gauge fields in $QCD$
are all $z$-independent, who live in the $z=0$ brane of
five-dimensional manifold.

The Lagrangian of a nonmass eigenstate ${\Phi}(x,z)$ of free
spin-$\frac{1}{2}$ fields is of the form\footnote{${\cal L}_{1n}$,
${\cal L}_{1m}$ denote the Lagrangian of one nonmass eigenstate or
one mass eigenstate respectively. ${\cal L}_{2n}$, ${\cal L}_{2m}$
have the similar meanings.}
\begin{eqnarray}\label{2150}
 {\cal L}_{1n}={\bar
{\Phi}}(x,z)\left(i\gamma^{\mu}{\partial}_{\mu}+i{\partial}_{z}\right)
\Phi(x,z)~,
\end{eqnarray}
here $\bar{\Phi}\equiv{\Phi}^{\dag}\gamma^{0}$ is called the spinor
adjoint to $\Phi$. I indicated that the mass eigenstate of a
spin-$\frac{1}{2}$ field satisfies $\Phi(x,z)=e^{imz}\phi(x)$ in
Ref.\cite{hua09a}, where $m$ is the rest mass. Therefore one can
obtain the Lagrangian of the mass eigenstate of a free
spin-$\frac{1}{2}$ field from (\ref{2150}), that is
\begin{eqnarray}\label{y2161}
 {\cal L}_{1m}={\bar
{\phi}}(x)\left(i\gamma^{\mu}{\partial}_{\mu}-m\right) \phi(x)~,
\end{eqnarray}
where $\bar{\phi}\equiv{\phi}^{\dag}\gamma^{0} $ is the spinor
adjoint to $\phi$.

Let's consider a quantum field system in which two different nonmass
eigenstates $\Psi_{1}(x,z)$ and $\Psi_{2}(x,z)$ of free
spin-$\frac{1}{2}$ fields form an isospin doublet as follows
\begin{eqnarray}
\label{y0002} \Psi(x,z)= \left(
\begin{array}{c}
\Psi_{1}(x,z)
\\
\Psi_{2}(x,z)
\end{array}
\right)~.
\end{eqnarray}
So the Lagrangian of two nonmass eigenstates of free
spin-$\frac{1}{2}$ fields is
\begin{eqnarray}\label{y2150}
 {\cal L}_{2n}={\bar
{\Psi}}(x,z)\left(i\gamma^{\mu}{\partial}_{\mu}+i{\partial}_{z}\right)
\Psi(x,z)~.
\end{eqnarray}

Here Let's first consider a special case: if $\Psi_{1}(x,z)$ and
$\Psi_{2}(x,z)$ are mass eigenstates with the same rest mass, then
$\Psi_{1}(x,z)=e^{imz}\psi_{1}(x)$, and
$\Psi_{2}(x,z)=e^{imz}\psi_{2}(x)$, where $m$ is the rest mass. I
can therefore obtain $\Psi(x,z)=e^{imz}\psi(x)$ by defining
\begin{eqnarray}
\label{y2156} \psi(x)= \left(
\begin{array}{c}
\psi_{1}(x)
\\
\psi_{2}(x)
\end{array}
\right)~.
\end{eqnarray}
Hence the Lagrangian of a quantum field system where two mass
eigenstates $e^{imz}\psi_{1}(x)$ and $e^{imz}\psi_{2}(x)$ form an
isospin doublet is acquired from (\ref{y2150}), namely
\begin{eqnarray}\label{y2160}
 {\cal L}_{2m}={\bar {\psi}}(x)\left(i\gamma^{\mu}{\partial}_{\mu}-m\right)
\psi(x)~.
\end{eqnarray}

The largest inner gauge symmetry group in this system is obviously
$SU(2)\times U(1)$. The total Lagrangian of this system reads
\begin{eqnarray}\nonumber
&& {\cal L}_{2mt}=i{\bar {\psi}}\gamma^{\mu}(\partial_{\mu}-i
g^{\prime} {\bf T}\cdot {\bf
B}_{\mu})\psi-e{\bar{\psi}}\gamma^{\mu}A_{\mu}\psi
\\
\label{y0001+} &&~~~~~~~-m{\bar {\psi}}\psi-\frac{1}{4}~{\tilde{\bf
F}}_{\mu\nu}\cdot{\tilde{\bf F}}^{\mu\nu}-~\frac{1}{4}~{\tilde
E}_{\mu\nu}{\tilde E}^{\mu\nu}~,
\end{eqnarray}
where $e$, $g^{\prime}$ are the coupling constants of $U(1)$ and
$SU(2)$ gauge fields respectively, and the dot ``$\cdot$" denotes a
scalar product in the isospace. In this case, ${\bf T}\cdot {\bf
B}_{\mu}$ means
\begin{eqnarray}\label{y2170}
 {\bf T}\cdot {\bf B}_{\mu}=T^{1}B^{1}_{\mu}+T^{2}B^{2}_{\mu}+T^{3}B^{3}_{\mu}~,
\end{eqnarray}
where $T^{a},~a=1,2,3$ are the generators of $SU(2)$ group, which
are written as
\begin{equation}
\label{y0008} T^{a}=\frac{1}{2}\tau^{a}~,
\end{equation}
where $\tau^{a}$ are the traceless matrices
\begin{eqnarray}
\label{y0006}\tau^{1}= \left(
\begin{array}{cc}
0 &  1
\\
1 & 0
\end{array}
\right)~,~~\tau^{2}= \left(
\begin{array}{cc}
0 &  -i
\\
i & 0
\end{array}
\right)~,~~\tau^{3}= \left(
\begin{array}{cc}
1 &  0
\\
0 & -1
\end{array}
\right)~,
\end{eqnarray}
known as the Pauli matrices. They obey the commutation relations
\begin{equation}
\label{y0010}
[\tau^{a},\tau^{b}]=2i\sum^{3}_{c=1}\varepsilon_{abc}\tau^{c}~.
\end{equation}
Here $\varepsilon_{abc}$ is the totally antisymmetry tensor in
3-dimensions.

In Yang-Mills theory \cite{yan54}, ${\bf T}\cdot {\bf B}_{\mu}$ is
called the $SU(2)$ gauge field, and its field strength tensor is of
the form
\begin{eqnarray}\label{y2180}
{\tilde{\bf F}}_{\mu\nu}\cdot{\bf T}=\partial_{\mu}({\bf B}_{\nu}
\cdot {\bf T} )-\partial_{\nu}({\bf B}_{\mu} \cdot {\bf T})-{ i}
g^{\prime}[{\bf B}_{\mu} \cdot {\bf T},{\bf B}_{\nu} \cdot {\bf
T}]~.
\end{eqnarray}
Hence the field strength ${\tilde{\bf F}}_{\mu\nu}$ satisfies
\begin{equation}
\label{y0004} {\tilde{\bf F}}_{\mu\nu}=\partial_{\mu}{\bf
B}_{\nu}-\partial_{\nu}{\bf B}_{\mu}+g^{\prime} {\bf B}_{\mu}\times
{\bf B}_{\nu} ~.
\end{equation}
I use $A_{\mu}$ to denote the $U(1)$ gauge field. The field strength
of the $U(1)$ gauge field is defined by
\begin{equation}
\label{y0005} {\tilde E}_{\mu\nu}=\partial_{\mu}
A_{\nu}-\partial_{\nu} A_{\mu}~.
\end{equation}
We are very familiar with above $SU(2)$ gauge fields and $U(1)$
gauge field which have been the fundamental content of quantum field
theories. The gauge bosons are massless. From a viewpoint of the
rest mass operator, above $SU(2)$ gauge fields couple to the mass
eigenstates with the same rest mass. I call this Yang-Mills fields
the first kind. The $SU(3)$ gauge fields in $QCD$ obviously belong
to this kind.

From now on I will study another kind of Yang-Mills fields
carefully. To make the gauge invariance explicit, let's formally
introduce the extra dimension $x^{4}$ as follows
\begin{equation}
\label{4302} x^{4}=-x_{4}=z~,~~~~\gamma^{4}=-\gamma_{4}={\bf 1}~.
\end{equation}
Hence the Lagrangian ${\cal L}_{2n}$ is rewritten as
\begin{eqnarray}\label{y2155}
 {\cal L}_{2n}=i{\bar
{\Psi}}(x,z)\gamma^{\alpha}{\partial}_{\alpha}
\Psi(x,z)~,~~~~\alpha=0,1,2,3,4~.
\end{eqnarray}
Since two different nonmass eigenstates $\Psi_{1}(x,z)$ and
$\Psi_{2}(x,z)$ form an isospin doublet, I can consider a local
isospin rotation logically similar to what Yang and Mills did in
their original paper. That is
\begin{eqnarray}\label{y5000}
{\Psi}^{\prime}(x,z)=S(x,z)\Psi(x,z)~,
\end{eqnarray}
where $S(x,z)$ is a $2\times2$ matrix. To make sure that the
probability density ${\bar {\Psi}}(x,z)\Psi(x,z)$ is invariant under
above rotation (\ref{y5000}), the matrix $S(x,z)$ must be unitary
with unit determinant
\begin{eqnarray}\label{y5003}
S^{\dag}(x,z)S(x,z)=1~.
\end{eqnarray}
All the matrices satisfy this condition generate the group $SU(2)$,
which is a non-Abelian Lie group. The transformation (\ref{y5000})
directly means that
\begin{eqnarray}\label{y5005}
{\bar {\Psi}}^{\prime}(x,z)={\bar {\Psi}}(x,z)S^{\dag}(x,z)~.
\end{eqnarray}
The matrix $S(x,z)$ can be written in the form
\begin{eqnarray}\label{y5010}
S(x,z)=\exp\left(i\sum_{a=1}^{3}\frac{{\tau}^{a}}{2}
{\Theta}^{a}(x,z)\right)~.
\end{eqnarray}
To discuss the gauge invariance, here I introduce the
gauge-invariant derivative
\begin{eqnarray}\label{y5015}
\hat{D}_{\alpha}=\partial_{\alpha}-ig_{1}{\bf T}\cdot{\bf
W}_{\alpha}(x,z)~,
\end{eqnarray}
where $g_{1}$ is the coupling constant of $SU(2)$ gauge fields, and
\begin{eqnarray}\label{y5020}
{\bf T}\cdot{\bf W}_{\alpha}(x,z)=\sum_{a=1}^{3}T^{a}{\bf
W}_{\alpha}^{a}(x,z)~.
\end{eqnarray}
Invariance requires that
\begin{eqnarray}\label{y5025}
(\partial_{\alpha}-ig_{1}{\bf T}\cdot{\bf
W}_{\alpha}^{\prime}){\Psi}^{\prime}=S(\partial_{\alpha}-ig_{1}{\bf
T}\cdot{\bf W}_{\alpha})\Psi~.
\end{eqnarray}
Combining (\ref{y5000}) and (\ref{y5025}), I obtain the gauge
transformation on ${\bf W}_{\alpha}$:
\begin{eqnarray}\label{y5030}
{\bf T}\cdot{\bf W}_{\alpha}^{\prime}=S{\bf T}\cdot{\bf
W}_{\alpha}S^{-1}+\frac{i}{g_{1}}S\left(\partial_{\alpha}S^{-1}\right)~.
\end{eqnarray}
In analogy to the procedure of obtaining gauge invariant field
strengths in electromagnetic case, I define now
\begin{eqnarray}\nonumber
&&{{\bf F}}_{\alpha\beta}\cdot{\bf T}=\sum_{a=1}^{3}{\bf
F}^{a}_{\alpha\beta}{T}^{a}=\hat{D}_{\alpha}({\bf T}\cdot{\bf
W}_{\beta})-\hat{D}_{\beta}({\bf T}\cdot{\bf W}_{\alpha})
\\
\nonumber&&~~~~~~~~~=\partial_{\alpha}({\bf W}_{\beta} \cdot {\bf T}
)-\partial_{\beta}({\bf W}_{\alpha} \cdot {\bf T})-{i}
g_{1}\left[{\bf W}_{\alpha} \cdot {\bf T},{\bf W}_{\beta} \cdot {\bf
T}\right]
\\\nonumber
&&~~~~~~~~~=(\partial_{\alpha}{\bf W}_{\beta}) \cdot {\bf T}
-(\partial_{\beta}{\bf W}_{\alpha}) \cdot {\bf T}+
g_{1}\sum_{abc}{\bf W}^{a}_{\alpha}{\bf W}^{b}_{\beta}
\varepsilon_{abc}T^{c}
\\\label{y2180+}
&&~~~~~~~~~=(\partial_{\alpha}{\bf W}_{\beta}-\partial_{\beta}{\bf
W}_{\alpha} + g_{1}{\bf W}_{\alpha}\times{\bf W}_{\beta} )\cdot {\bf
T} ~.
\end{eqnarray}
Therefore the isovector of field strengths is
\begin{eqnarray}\label{y5035}
{{\bf F}}_{\alpha\beta}=\partial_{\alpha}{\bf
W}_{\beta}-\partial_{\beta}{\bf W}_{\alpha} + g_{1}{\bf
W}_{\alpha}\times{\bf W}_{\beta} ~.
\end{eqnarray}
One easily shows from the equation (\ref{y5030}) that
\begin{eqnarray}\label{y5040}
{\bf F}^{\prime}_{\alpha\beta}\cdot{\bf T}=S{{\bf
F}}_{\alpha\beta}\cdot{\bf T}S^{-1} ~.
\end{eqnarray}
I obtain a gauge invariant Lagrangian by performing the trace over
the isospin indices:
\begin{eqnarray}\nonumber
&&{\cal L}_{SU(2)}= -\frac{1}{2}{\rm Tr} \{({{\bf
F}}_{\alpha\beta}\cdot{\bf T})({{\bf F}}^{\alpha\beta}\cdot{\bf T})
\} \\
\label{y5045} &&~~~~~~~~=-\frac{1}{4}{{\bf
F}}_{\alpha\beta}\cdot{{\bf
F}}^{\alpha\beta}=-\frac{1}{4}\sum_{a=1}^{3}{{\bf
F}}^{a}_{\alpha\beta}{{\bf F}}^{a\alpha\beta} ~.
\end{eqnarray}
Considering the couplings between fermions and gauge bosons and the
self-couplings of gauge bosons, one can get the complete Lagrangian
as follows
\begin{eqnarray}\nonumber
&&{\cal L}_{2nt}={\cal L}_{2n}+{\cal L}_{int}+{\cal L}_{SU(2)}
\\\label{y5050}
&&~~~~~~=i{\bar {\Psi}}\gamma^{\alpha}(\partial_{\alpha}-ig_{1}{\bf
T}\cdot{\bf W}_{\alpha}) \Psi-\frac{1}{4}{{\bf
F}}_{\alpha\beta}\cdot{{\bf F}}^{\alpha\beta}~.
\end{eqnarray}

In order to build a foundation for setting up an electroweak model
without Higgs mechanism, I discuss the $SU(2)\times U(1)$ gauge
fields in this letter. The $U(1)$ gauge field that couples to a
nonmass eigenstate has been studied in my preceding
paper~\cite{hua09a}. The Lagrangian (\ref{y2155}) shows me that the
maximal gauge groups for this quantum field system are $SU(2)\times
U(1)$. I have introduced the $SU(2)$ gauge fields in this system,
now I put in the $U(1)$ gauge field. Let us multiply the nonmass
eigenstates $\Psi(x,z)$ by a local phase $e^{i\Theta(x,z)}$, namely
\begin{equation}
\label{y5055} \Psi^{\prime\prime}=~e^{i\Theta(x,z)}\Psi~,~{\bar
\Psi}^{\prime\prime}=~e^{-i\Theta(x,z)}{\bar \Psi}~.
\end{equation}
According to the discussion in Ref.\cite{hua09a}, I introduce the
 $U(1)$ gauge field of the second kind, that is ${\bf X}_{\alpha}(x,z)$. Under the
transformation of (\ref{y5055}), ${\bf X}_{\alpha}(x,z)$ transforms
as
\begin{equation}
\label{y5060} {\bf X}^{\prime\prime}_{\alpha}(x,z)={\bf
X}_{\alpha}(x,z)+ \frac{1}{g_{2}}{\partial_{\alpha}}\Theta(x,z)~,
\end{equation}
here $g_{2}$ is the coupling constant of $U(1)$. The strength tensor
of $U(1)$ gauge field is of the form
\begin{equation}
\label{y5065} {\bf E}_{\alpha\beta}(x,z)={\partial}_{{\alpha}} {\bf
X}_{\beta}(x,z)-{\partial}_{\beta} {\bf X}_{\alpha}(x,z)~,
\end{equation}
which is invariant under the transformations of (\ref{y5055}) and
(\ref{y5060}). Therefore the total Lagrangian including the $U(1)$
gauge field of the second kind is written as
\begin{eqnarray}\nonumber
&&{\cal L}_{total}={\cal L}_{2n}+{\cal L}_{int}+{\cal
L}_{U(1)}+{\cal L}_{SU(2)}
\\\nonumber
&&~~~~~~~=i{\bar {\Psi}}\gamma^{\alpha}(\partial_{\alpha}-ig_{2}{\bf
X}_{\alpha}-ig_{1}{\bf T}\cdot{\bf W}_{\alpha}) \Psi
\\\label{y5070}
&&~~~~~~~~~~-\frac{1}{4}{{\bf E}}_{\alpha\beta}{\bf
E}^{\alpha\beta}-\frac{1}{4}{{\bf F}}_{\alpha\beta}\cdot{{\bf
F}}^{\alpha\beta}~.
\end{eqnarray}
Obviously the total Lagrangian is also invariant under the
transformations of (\ref{y5055}) and (\ref{y5060}). The gauge
covariance requires that ${\bf X}_{\alpha}$ and ${\bf T}\cdot{\bf
W}_{\alpha}$ are all nonmass eigenstates.

Till now I merely constructed the $SU(2)\times U(1)$ gauge fields on
a five-dimensional manifold, which is quite the same as the gauge
fields proposed by Yang and Mills. Yes, from the five-dimensional
point of view, the gauge bosons are massless in above discussed
$SU(2)\times U(1)$ gauge fields since there are no mass term in the
total Lagrangian (\ref{y5070}). But, things will be quite different
when I discuss them from the viewpoint of $z=0$ brane.

I have pointed out that the gravitational field, the electromagnetic
field and $SU(3)$ gauge fields in $QCD$ are living in the $z=0$
brane of five-dimensional manifold. Also I have proved that the
$z$-independent electromagnetic field, gravitational field and
$SU(3)$ gauge fields only couple to the mass eigenstates. Therefore
I can find that the mass eigenstates coupled by the gravitation, the
electromagnetic field and the gluon fields are also living in the
$z=0$ 4-dimensional brane.

It is indicated that the nonmass eigenstate is composed of mass
eigenstates~\cite{hua09a}. To discuss the physical properties of the
mass eigenstates who compose the gauge fields ${\bf X}_{\alpha}$ and
${\bf W}_{\alpha}$, I write out the spacetime component and
$z$-related component of gauge fields separately. Therefore
\begin{eqnarray}
\label{y5080} && {\bf X}_{\alpha}(x,z)\equiv ({\bf
X}_{\mu}(x,z),{\bf X}_{z}(x,z))~,
\\
\label{y5075} &&{\bf W}_{\alpha}(x,z)\equiv ({\bf W}_{\mu}(x,z),{\bf
W}_{z}(x,z))~.
\end{eqnarray}
To list the components of ${\bf W}_{\alpha}(x,z)$ manifestly,  I
rewrite (\ref{y5075}) as
\begin{equation}
\label{y5078} {\bf W}^{a}_{\alpha}(x,z)\equiv ({\bf
W}^{a}_{\mu}(x,z),{\bf W}^{a}_{z}(x,z))~,~a=1,2,3~.
\end{equation}
After that, the strength tensor ${\bf E}_{\alpha\beta}(x,z)$ is
correspondingly divided into three parts
\begin{eqnarray}
\label{y5089} && {\bf E}_{\mu\nu}(x,z)=\partial_{\mu} {\bf
X}_{\nu}(x,z)-\partial_{\nu} {\bf X}_{\mu}(x,z)~,
\\
\label{y5090} && {\bf E}_{\mu z}(x,z)=-{\bf
E}_{z\mu}(x,z)=\partial_{\mu} {\bf X}_{z}(x,z)-\partial_{z} {\bf
X}_{\mu}(x,z)~,
\\
\label{y5091+} && {\bf E}_{zz}(x,z)=\partial_{z} {\bf
X}_{z}(x,z)-\partial_{z} {\bf X}_{z}(x,z)\equiv 0~.
\end{eqnarray}
Surely one can also get the decomposition of the strength tensor
${\bf F}_{\alpha\beta}(x,z)$ as follows
\begin{eqnarray}
\label{y5084} && {{\bf F}}_{\mu\nu}(x,z)=\partial_{\mu}{\bf
W}_{\nu}-\partial_{\nu}{\bf W}_{\mu}+g_{1} {\bf W}_{\mu}\times {\bf
W}_{\nu}~,
\\
\label{y5085} && {\bf F}_{\mu z}(x,z)=-{\bf
F}_{z\mu}(x,z)=\partial_{\mu}{\bf W}_{z} -\partial_{z}{\bf W}_{\mu}+
g_{1}{\bf W}_{\mu} \times {\bf W}_{z}~,
\\
\label{y5086} && {\bf F}_{zz}(x,z)\equiv 0~.
\end{eqnarray}

Now let's consider the movement of gauge bosons. Firstly, the
interaction term ${\cal L}_{int}$ in total Lagrangian (\ref{y5070})
shows that the movement of gauge bosons is decided by the momentum
of $\Psi$ and $\bar{\Psi}$. The $\Psi$ and $\bar{\Psi}$ are nonmass
eigenstates, who are composed of mass eigenstates that are living in
the $z=0$ brane and moving along the $z=0$ brane, hence gauge bosons
must move along the $z=0$ brane. Secondly, once the gauge bosons are
produced, they are constrained by gravitation, which is living in
the $z=0$ brane. Consequently
\begin{equation}
\label{y5120} {\bf W}_{z}(x,z)=0~,~{\bf X}_{z}(x,z)=0~.
\end{equation}
Hence the equations (\ref{y5090}) and (\ref{y5085}) reduce to
\begin{eqnarray}
\label{y5130} &&{\bf E}_{\mu z}(x,z)=-{\bf
E}_{z\mu}(x,z)=-\partial_{z} {\bf X}_{\mu}(x,z)~,
\\
\label{y5125} &&{\bf F}_{\mu z}(x,z)=-{\bf
F}_{z\mu}(x,z)=-\partial_{z}{\bf W}_{\mu}(x,z)~.
\end{eqnarray}
The spacetime components ${\bf X}_{\mu}(x,z)$ and ${\bf
W}_{\mu}(x,z)$ are nonmass eigenstates, which are linear
combinations of mass eigenstates. I define the mass eigenstates of
bosons $W^{\pm}$ by
\begin{eqnarray}\label{y5101}
&&{\bf W}^{+}_{\mu}(x,z)=\frac{1}{\sqrt{2}}\left({\bf
W}^{1}_{\mu}(x,z)-i{\bf W}^{2}_{\mu}(x,z)\right)~,
\\\label{y5100}
&&{\bf W}^{-}_{\mu}(x,z)=\frac{1}{\sqrt{2}}\left({\bf
W}^{1}_{\mu}(x,z)+i{\bf W}^{2}_{\mu}(x,z)\right)~.
\end{eqnarray}
When the mass eigenstates of $W^{\pm}$ are expressed by
\begin{eqnarray}\label{y5102}
{\bf W}^{+}_{\mu}(x,z)=e^{im_{W}z}{W}^{+}_{\mu}(x)~,~{\bf
W}^{-}_{\mu}(x,z)=e^{im_{W}z}{W}^{-}_{\mu}(x)~,
\end{eqnarray}
$m_{W}$ being the rest mass of $W^{\pm}$, the nonmass eigenstates
${\bf W}^{1}_{\mu}(x,z)$ and ${\bf W}^{2}_{\mu}(x,z)$ are manifestly
given by
\begin{eqnarray}\nonumber
&&{\bf W}^{1}_{\mu}(x,z)=\frac{1}{\sqrt{2}}\left({\bf
W}^{+}_{\mu}(x,z)+{\bf W}^{-}_{\mu}(x,z)\right)
\\\label{y5105}
&&~~~~~~~~~~~=\frac{1}{\sqrt{2}}e^{im_{W}z}\left({W}^{+}_{\mu}(x)+{
W}^{-}_{\mu}(x)\right)~,
\end{eqnarray}
and
\begin{eqnarray}\nonumber
&&{\bf W}^{2}_{\mu}(x,z)=\frac{i}{\sqrt{2}}\left({\bf
W}^{+}_{\mu}(x,z)-{\bf W}^{-}_{\mu}(x,z)\right)
\\\label{y5110}
&&~~~~~~~~~~~=\frac{i}{\sqrt{2}}e^{im_{W}z}\left({W}^{+}_{\mu}(x)-{
W}^{-}_{\mu}(x)\right)~.
\end{eqnarray}
The boson fields ${\bf W}^{+}_{\mu}(x,z)$, ${\bf W}^{-}_{\mu}(x,z)$,
${\bf Z}_{\mu}(x,z)$ and photon field $A_{\mu}(x)$, which are mass
eigenstates, constitute a complete Hilbert space. From
Ref.\cite{hua09a}, I know that
\begin{eqnarray}\label{y5104}
{\bf Z}_{\mu}(x,z)=e^{im_{Z}z}Z_{\mu}(x)~,
\end{eqnarray}
here $m_{Z}$ is the rest mass of boson $Z^{0}$. The nonmass
eigenstates ${\bf W}^{3}_{\mu}(x,z)$ and ${\bf X}_{\mu}(x,z)$ are
the linear combinations of ${\bf Z}_{\mu}(x,z)$ and $A_{\mu}(x)$,
namely
\begin{eqnarray}\nonumber
&&{\bf W}^{3}_{\mu}(x,z)=\sin \theta_{W}A_{\mu}+\cos\theta_{W}{\bf
Z}_{\mu}(x,z)
\\\label{y5091}
&&~~~~~~~~~~~=\sin
\theta_{W}A_{\mu}+\cos\theta_{W}e^{im_{Z}z}Z_{\mu}~,
\\
\nonumber
&&{\bf X}_{\mu}(x,z)=\cos \theta_{W}A_{\mu}-\sin\theta_{W}{\bf
Z}_{\mu}(x,z)
\\\label{y5095}
&&~~~~~~~~~~~=\cos
\theta_{W}A_{\mu}-\sin\theta_{W}e^{im_{Z}z}Z_{\mu}~,
\end{eqnarray}
where $\theta_{W}$ is the Weinberg angle.

The nonmass eigenstates ${\bf W}_{\alpha}^{a} (a=1,2,3)$ in ${\bf
T}\cdot{\bf W}_{\alpha}$ must have the same rest mass because of two
reasons: Each ${\bf W}_{\alpha}^{a}$ plays the quite equal role in
gauge field ${\bf T}\cdot{\bf W}_{\alpha}$; The model must be
$SU(2)$ gauge invariant. Consequently
\begin{eqnarray}\label{y5265}
m_{{\bf W}^{1}_{\mu}}=m_{{\bf W}^{2}_{\mu}}=m_{{\bf W}^{3}_{\mu}}~.
\end{eqnarray}
In Ref.\cite{hua09a}, it is indicated that the rest mass squared of
nonmass eigenstate of vector fields can be calculated, that is
\begin{eqnarray}\label{y5300}
m_{{\bf V}_{\mu}}^{2}=\sum_{j=1}^{n}a_{j}a^{*}_{j} m_{j}^{2}
\end{eqnarray}
is right if and only if
\begin{eqnarray}\label{y5305}
{\bf V}_{\mu}=\sum_{j=1}^{n}a_{j}[{\bf
V}_{\mu}]_{j}=\sum_{j=1}^{n}a_{j}e^{i m_{j} z}[V_{\mu}]_{j}~.
\end{eqnarray}
Then one can easily obtain the rest masses of ${\bf W}^{1}_{\mu}$
and ${\bf W}^{2}_{\mu}$ from (\ref{y5105}) and (\ref{y5110})
respectively
\begin{eqnarray}\label{y5270}
m_{{\bf W}^{1}_{\mu}}^{2}=m_{{\bf W}^{2}_{\mu}}^{2}=m_{W}^{2}~,
\end{eqnarray}
also get the rest mass of ${\bf W}^{3}_{\mu}$ from (\ref{y5091})
\begin{eqnarray}\label{y5275}
m_{{\bf W}^{3}_{\mu}}^{2}=m_{Z}^{2}(\cos\theta_{W})^2~.
\end{eqnarray}
Therefore combining (\ref{y5265}), (\ref{y5270}) and (\ref{y5275}),
I obtain the following relation
\begin{eqnarray}\label{y5280}
m_{W}=m_{Z}\cos\theta_{W}~.
\end{eqnarray}

In the total Lagrangian (\ref{y5070}), the kinetic terms of Fermions
and the interaction terms will be discussed carefully in my
forthcoming paper~\cite{hua09c}, in this letter I only discuss the
self-couplings of $SU(2)\times U(1)$ gauge fields, namely the terms
${\cal L}_{U(1)}+{\cal L}_{SU(2)}$ in (\ref{y5070}). It has been
pointed out that the gauge bosons in my model merely propagate along
the $z=0$ brane, therefore ${\bf W}_{z}(x,z)=0,~{\bf X}_{z}(x,z)=0$.
In this case, substituting (\ref{y5095}) into (\ref{y5130}) yields
\begin{eqnarray}\label{y5150}
&&{{\bf E}}_{z\mu}=-im_{Z}\sin\theta_{W}e^{im_{Z}z}Z_{\mu}~.
\end{eqnarray}
Substituting (\ref{y5105}), (\ref{y5110}) and (\ref{y5091}) into
(\ref{y5125}) respectively, I obtain
\begin{eqnarray}
\label{y5151} &&{{\bf
F}}_{z\mu}^{1}=\frac{i}{\sqrt{2}}m_{W}e^{im_{W}z}\left({W}^{+}_{\mu}+{
W}^{-}_{\mu}\right)~,
\\\label{y5152}
&&{{\bf
F}}_{z\mu}^{2}=-\frac{1}{\sqrt{2}}m_{W}e^{im_{W}z}\left({W}^{+}_{\mu}-{
W}^{-}_{\mu}\right)~,
\\\label{y5153}
&&{{\bf F}}_{z\mu}^{3}=im_{Z}\cos\theta_{W}e^{im_{Z}z}Z_{\mu}~.
\end{eqnarray}
Substituting (\ref{y5150}), (\ref{y5151}), (\ref{y5152}) and
(\ref{y5153}) into ${\cal L}_{U(1)}+{\cal L}_{SU(2)}$, I find that
the self-coupling terms of $SU(2)\times U(1)$ gauge fields become
\begin{eqnarray}\nonumber
&&~~~~{\cal L}_{U(1)}+{\cal L}_{SU(2)}
\\\nonumber
&&=-\frac{1}{4}{{\bf E}}_{\alpha\beta}{\bf
E}^{\alpha\beta}-\frac{1}{4}{{\bf F}}_{\alpha\beta}\cdot{{\bf
F}}^{\alpha\beta}
\\\nonumber
&&=-\frac{1}{4}{{\bf E}}_{\mu\nu}{\bf E}^{\mu\nu}-\frac{1}{4}{{\bf
F}}_{\mu\nu}\cdot{{\bf F}}^{\mu\nu}-\frac{1}{2}\left({{\bf
E}}_{z\mu}{\bf E}^{z\mu}+{{\bf F}}_{z\mu}\cdot{{\bf
F}}^{z\mu}\right)
\\\nonumber
&&=-\frac{1}{4}{{\bf E}}_{\mu\nu}{\bf E}^{\mu\nu}-\frac{1}{4}{{\bf
F}}_{\mu\nu}\cdot{{\bf F}}^{\mu\nu}-\frac{1}{2}\left({{\bf
E}}_{z\mu}{\bf E}^{z\mu}+{\bf F}^{1}_{z\mu}{{\bf F}}^{1z\mu}+{\bf
F}^{2}_{z\mu}{{\bf F}}^{2z\mu}+{\bf F}^{3}_{z\mu}{{\bf
F}}^{3z\mu}\right)
\\\label{y5135}
&&=-\frac{1}{4}{{\bf E}}_{\mu\nu}{\bf E}^{\mu\nu}-\frac{1}{4}{{\bf
F}}_{\mu\nu}\cdot{{\bf
F}}^{\mu\nu}+\frac{1}{2}m_{Z}^{2}e^{2im_{Z}z}Z_{\mu}Z^{\mu}+m_{W}^{2}e^{2im_{W}z}W_{\mu}^{+}W^{\mu
-}~.
\end{eqnarray}

It is indicated that all the mass eigenstates coupled by the
gravitation, the electromagnetic field and the gluon fields are
living in the $z=0$ brane. Hence, expressed by the mass eigenstates
in the $z=0$ brane, the self-coupling terms of $SU(2)\times U(1)$
gauge fields reduce to
\begin{eqnarray}\nonumber
\nonumber &&~~~~{\cal L}_{U(1),z=0}+{\cal L}_{SU(2),z=0}
\\\label{y5140}
&&=-\frac{1}{4}{E}_{\mu\nu}{E}^{\mu\nu}-\frac{1}{4}{F}_{\mu\nu}\cdot{F}^{\mu\nu}
+\frac{1}{2}m_{Z}^{2}Z_{\mu}Z^{\mu}+m_{W}^{2}W_{\mu}^{+}W^{\mu -}~,
\end{eqnarray}
where
${F}_{\mu\nu}\equiv\{{F}^{1}_{\mu\nu},{F}^{2}_{\mu\nu},{F}^{3}_{\mu\nu}\}$,
and ${E}_{\mu\nu}$ is formulated by
\begin{eqnarray}
\label{y5285} {E}_{\mu\nu}(x)=\partial_{\mu} {
X}_{\nu}(x)-\partial_{\nu} {X}_{\mu}(x)~,
\end{eqnarray}
and ${F}_{\mu\nu}$ is given by
\begin{eqnarray}
\label{y5290} {F}_{\mu\nu}(x)=\partial_{\mu}{
W}_{\nu}(x)-\partial_{\nu}{W}_{\mu}(x)+g_{1} {W}_{\mu}(x)\times
{W}_{\nu}(x)~,
\end{eqnarray}
in which
${W}_{\mu}(x)\equiv\{{W}^{1}_{\mu}(x),{W}^{2}_{\mu}(x),{W}^{3}_{\mu}(x)\}$.
The four-dimensional fields ${X}_{\mu}(x)$ and ${W}_{\mu}(x)$ are
composed of mass eigenstates which are constrained in the $z=0$
brane. From (\ref{y5105}), (\ref{y5110}), (\ref{y5091}) and
(\ref{y5095}), one can easily obtain the expressions of them in the
following
\begin{eqnarray}\label{y5245}
&&{W}^{1}_{\mu}(x)=\frac{1}{\sqrt{2}}\left({W}^{+}_{\mu}(x)+{
W}^{-}_{\mu}(x)\right)~, \\\label{y5250}
&&{W}^{2}_{\mu}(x)=\frac{i}{\sqrt{2}}\left({W}^{+}_{\mu}(x)-{
W}^{-}_{\mu}(x)\right)~, \\\label{y5255} &&{W}^{3}_{\mu}(x)=\sin
\theta_{W}A_{\mu}(x)+\cos\theta_{W}Z_{\mu}(x)~,
\\\label{y5260}
&&{X}_{\mu}(x)=\cos \theta_{W}A_{\mu}(x)-\sin\theta_{W}Z_{\mu}(x)~.
\end{eqnarray}
Obviously they have the same forms as the definitions of gauge
bosons of $SU(2)\times U(1)$ gauge fields in the
Glashow-Salam-Weinberg model~\cite{gm00}. The fields
${W}^{+}_{\mu}(x)$, ${W}^{-}_{\mu}(x)$, $Z_{\mu}(x)$ and
$A_{\mu}(x)$ in above expressions are mass eigenstates that are
constrained in the $z=0$ brane.

The $SU(2)\times U(1)$ gauge fields of the second kind merely couple
to the nonmass eigenstates, which are the nonmass eigenstates as
well, hence cannot be observed directly. The nonmass eigenstates of
gauge fields are composed of the mass eigenstates that are
constrained in the $z=0$ brane. When I reexpress the Lagrangian
${\cal L}_{U(1)}+{\cal L}_{SU(2)}$ by the mass eigenstates of gauge
bosons who live in four-dimensional spacetime, I find that the
fields ${W}^{+}_{\mu}(x)$, ${W}^{-}_{\mu}(x)$ and $Z_{\mu}(x)$ can
be treated as massive gauge bosons from the four-dimensional point
of view since their mass terms automatically appear in the
four-dimensional Lagrangian (\ref{y5140}).

The mass terms of gauge bosons who are living in four-dimensional
spacetime aren't inserted by hands, which is produced automatically.
From five-dimensional point of view, the $SU(2)\times U(1)$ gauge
fields of the second kind are massless, which are the usual gauge
fields that we are very familiar with, since there are no mass term
in the total Lagrangian (\ref{y5070}). Therefore, the gauge fields
in this model should be renormalizable.

let me explicitly explain this model again: The general equation of
a nonmass eigenstate of spin-$\frac{1}{2}$ fields is built on a
five-dimensional manifold, therefore the gauge fields of the second
kinds, who couple to the nonmass eigenstates of spin-$\frac{1}{2}$
fields, are constructed on a five-dimensional manifold as well. But
the nonmass eigenstates are composed of mass eigenstates, which are
physically observable. The mass eigenstates are merely coupled by
the electromagnetic field, the gravitation and the gluon fields, who
are living in the $z=0$ brane. Hence the initial momentum of the
nonmass eigenstates of spin-$\frac{1}{2}$ fields are along the $z=0$
brane, which decides that the gauge fields of the second kind must
propagate along the $z=0$ spacetime as well. When the gauge fields
of the second kind are reexpressed by their mass eigenstates that
living in the four-dimensional spacetime, it is found that the gauge
bosons who are mass eigenstates can be treated as massive vector
fields.


\begin{thebibliography}{99}

\bibitem{yan54} C. N. Yang and R. L. Mills, Phys. Rev. {\bf 96}, 191 (1954).

\bibitem{hig64} P. W. Higgs,
Phys. Rev. Lett. {\bf 13}, 508 (1964); P. W. Higgs, Phys. Rev.
{\bf 145}, 1156 (1966); F. Englert and R. Brout, Phys. Rev. Lett.
{\bf 13}, 321 (1964).

\bibitem{gsw60}
S.~L.~Glashow, Nucl. Phys. {\bf 22}, 579 (1960); J.~Goldstone,
A.~Salam and S.~Weinberg, Phys.~Rev. {\bf 127}, 965 (1962);
S.~Weinberg, Phys.~Rev.~Lett. {\bf 19}, 1264 (1967);
S.~L.~Glashow, J.~Iliopoulos and L.~Maiani, Phys.~Rev.~D {\bf 2},
1285 (1970).



\bibitem{xin04} For recent reviews on neutrino masses and mixing angles, see Z. Z. Xing,
Int. J. Mod. Phys. A {\bf 19}, 1 (2004); R. D. McKeown and P.
Vogel, Phys. Rept. {\bf 394}, 315 (2004); M. C. Gonzalez-Garcia
and Y. Nir, Rev. Mod. Phys. {\bf 75}, 345 (2003); M.-C. Chen and
K. T. Mahanthappa, Int. J. Mod. Phys. A {\bf 18}, 5819 (2003).

\bibitem{hua09a} X.-B. Huang, ``{\em Nonmass Eigenstates of Boson and Fermion
Fields}", [arXiv:hep-th/0906.2441].

\bibitem{hua09c} X.-B. Huang, ``{\em An Electroweak Model without Higgs Mechanism}", in preparation.

\bibitem{gm00} W.~Greiner and B.~M\"{u}ller, {\em Gauge Theory of Weak Interactions},
(3rd. edition), (Springer-Verlag, 2000).



\end{thebibliography}
\end{document}